\documentclass[preprint,12pt]{elsarticle}

\usepackage{amssymb}
\usepackage{amsmath}

\usepackage{tikz}
\usepackage{amsmath}
\usepackage{ifthen}
\usepackage{tikz}
\usepackage{amsmath,amssymb}
\usepackage{colortbl}
\usepackage{xcolor}
\usepackage{tcolorbox}
\usepackage{verbatim}
\usepackage{listings}
\usepackage{url}
\usepackage{xspace}
\usepackage{ifthen}
\usepackage[T1]{fontenc}
\usepackage[inline]{enumitem}
\usepackage{etoolbox}

\AtBeginEnvironment{table}{\small}
\AtBeginEnvironment{tabular}{\small}

\usepackage[framemethod=tikz]{mdframed}
\usepackage{booktabs}   %% for \toprule / \midrule / \bottomrule in tables
\usepackage{caption}
\usepackage{booktabs}
\usepackage{tabularx}
\usepackage{booktabs}
\usepackage{tabularx}
\usepackage{array}
\newcolumntype{Y}[1]{>{\raggedright\arraybackslash\hsize=#1\hsize}X}
\usepackage[section]{placeins}
\journal{Computers and Security}
\newcommand{\toolName}{TeleGapper\xspace}

\newboolean{showcomments}
\setboolean{showcomments}{false}

\ifthenelse{\boolean{showcomments}}
  {\newcommand{\nb}[2]{
   \fbox{\bfseries\sffamily\scriptsize#1}
   {\sf\small$\blacktriangleright$\textit{\textcolor{red}{#2}}$\blacktriangleleft$}
  }}
  {\newcommand{\nb}[2]{}
   
  }

\begin{document}

\begin{frontmatter}



\title{TeleGapper: On the (un)reliability of Privacy Policies in Telegram Mini apps}


\author[1]{Luca Ferrari} %% Author name
\ead{luca.ferrari@imtlucca.it}

\affiliation[1]{
  organization={IMT School for Advanced Studies Lucca},
  addressline={Piazza San Francesco 19},
  city={Lucca},
  postcode={55100},
  country={Italy}
}

\author[2]{Mariano Ceccato}
\ead{mariano.ceccato@univr.it}

\affiliation[2]{
  organization={University of Verona},
  addressline={Strada le Grazie 15},
  city={Verona},
  postcode={37134},
  country={Italy}
}
%% --- Author 2 ---
\author[3]{Luca Verderame}
\ead{luca.verderame@unige.it}

\affiliation[3]{
  organization={University of Genova},
  addressline={Via Dodecaneso 35},
  city={Genova},
  postcode={16146},
  country={Italy}
}

%% Abstract
\begin{abstract}

Telegram Mini Apps are Web applications embedded within the Telegram client, forming an ecosystem of third-party services within one of the world's most widely used messaging platforms. Despite their growing adoption and access to Telegram-provided context, their privacy properties remain largely unexplored. Unlike ecosystems such as WeChat, which rely on tightly controlled, proprietary execution frameworks, Telegram adopts a different model: Mini Apps run inside a WebView, combining platform-provided context with standard Web capabilities and unrestricted outbound networking. This enables applications to transmit sensitive information to analytics, advertising, tracking, or other third parties through ordinary Web requests, often with limited visibility.

Privacy disclosures are therefore critical for transparency. Telegram allows Mini Apps either to define an application-specific privacy policy or to rely on a platform-provided default policy. While the latter reduces the developer's disclosure burden, it may lead to generic statements that do not accurately capture actual data practices of individual Mini Apps.

In this paper, we present TeleGapper, a black-box dynamic analysis framework to assess the privacy posture of Mini Apps by capturing runtime network traffic, identifying third-party communications, and comparing observed data flows against disclosed privacy information. We evaluate 278 working Mini Apps collected from tApps Center, a community-driven catalogue for discovering third-party applications in Telegram. We find that 59.4\% contact at least one undisclosed third party, 78.8\% rely exclusively on Telegram's default privacy policy, and none provides a consent or opt-out mechanism. These findings expose a substantial transparency and compliance gap in a widely used yet understudied ecosystem.

\end{abstract}


% \begin{highlights}
% \item First black box analysis tool for Telegram Mini App
% \item 59.4\% of 278 Mini Apps contact third parties their policy never discloses
% \item 78.8\% rely on Telegram's default policy; custom notices are no more accurate
% \item 85.5\% of violating apps contact undisclosed third parties at page load
% \item None of the 278 apps offered any consent banner, dialog, or opt-out
% \end{highlights}

\begin{highlights}
\item First black-box dynamic analysis framework for Telegram Mini Apps
\item 59.4\% of 278 Mini Apps contact at least one third party not disclosed by the applicable privacy policy
\item 78.8\% rely on Telegram’s default policy; custom policies show no statistically significant improvement in compliance
\item 85.5\% of violating apps contact undeclared third parties already at page load
\item None of the 278 apps presented an observable consent or opt-out mechanism at launch
\end{highlights}

%% Keywords
\begin{keyword}
Mini App \sep Telegram \sep Dynamic Analysis \sep Privacy Violation \sep Network Analysis

\end{keyword}

\end{frontmatter}

\section{Introduction}
\label{sec:intro}
\noindent

Mini apps have recently emerged as a rapidly expanding paradigm in mobile computing, consisting of small, lightweight software programs that run inside a larger “host” or “super app.” They support a broad range of services traditionally offered by standalone mobile applications, including e-commerce, food delivery, transportation, payments and financial services, entertainment, ticketing, and productivity tools, while providing users with direct access within the host platform. Notably, some of the most prominent mini app ecosystems are embedded in large-scale social and messaging platforms, such as WeChat, AliPay,  TikTok, and Telegram. By integrating third-party services directly into these widely used environments, mini apps allow users to access diverse functionality without leaving the host platform.

The tight integration with the super app, however, is also what makes mini apps privacy-sensitive: they execute within an already authenticated user session and can access account identity and device context through privileged host APIs.

Among these ecosystems, Telegram stands out both for its scale and for its distinct architecture. Its Mini Apps reached an estimated 150--190 million active users in 2025~\cite{telegramMiniAppActiveUsers}, yet have received comparatively little research attention. Unlike ecosystems such as WeChat, Alipay, and Baidu, where mini apps are packaged using platform-specific languages and manifests, submitted to the host platform, and distributed through its infrastructure, Telegram Mini Apps (hereafter, Mini Apps) are ordinary Web applications loaded from developer-controlled URLs inside a WebView. This differs substantially from the more standardized and tightly specified models adopted in other ecosystems, as Telegram retains the flexibility and openness of the Web.

This distinction is methodological as much as architectural. Prior analyses of unintended data collection in Chinese mini app ecosystems can inspect application packages before execution~\cite{wang2024minichecker, xiang2026minieval}. Telegram provides no equivalent package to decompile or manifest to audit: the application is retrieved dynamically from the developer's infrastructure at runtime. Consequently, static and taint-analysis approaches designed around inspectable mini app artifacts~\cite{wang2023taintmini, li2023minitracker} do not directly transfer, motivating a runtime perspective on the privacy behavior of Mini Apps.

Moreover, the Terms of Service for Mini Apps~\cite{TermsMiniApp} enumerate what a Mini App obtains the moment it is opened, without asking and before the user can act: the IP address, which follows from the page being fetched over HTTPS, and, through \texttt{initData}, the Telegram user ID, public name, username, profile picture, language tag, and premium status. Telegram, however, disclaims control over what is exchanged once these data have been transmitted to the Mini App.

Two consequences follow. First, a Mini App does not need to request an identity explicitly: it receives one before the user can perform any action, meaning that data collection can begin as early as Mini App load. Second, once that identity has been disclosed, the privacy policy becomes the primary means through which users can understand how it is subsequently processed and shared: a custom notice where the developer publishes one, or Telegram's Standard Bot Privacy Policy~\cite{MiniAppPP}, which applies by default otherwise.

This arrangement is governed by privacy regulations that place obligations on the developer, who typically acts as the controller for the data processed by the Mini App. Articles~13 and~14 GDPR~\cite{art13,art14} require the recipients of personal data, or at least their categories, to be disclosed to the data subject. Both are relevant here: the Mini App may collect data directly from the user while receiving other data from Telegram. Article~5(3) of the ePrivacy Directive~\cite{eprivacy} further requires prior consent for storing information in, or accessing information from, the user's terminal equipment, unless a statutory exception applies.

These conditions create an accountability gap between declared and actual data practices. While similar risks arise across super-app ecosystems, which centralize user data and delegate access to mini apps~\cite{yang2025understanding}, verifying whether the data flows generated by an individual mini app are consistent with its privacy disclosures remains challenging. The developer is the only party that directly knows both the recipients declared in the policy and those actually contacted; users cannot readily observe where their data are forwarded, and external auditors or data protection authorities cannot realistically inspect every application individually. This problem is particularly acute for Telegram: its Web-based architecture provides no publicly inspectable application artifact from which such flows can be inferred before execution. As a result, there is currently no systematic way for external parties to determine which third parties receive Mini App data, whether those recipients are disclosed, and when such transfers occur.

To investigate this gap in the Telegram ecosystem, we define four research questions (RQs):
\begin{itemize}
\item \textbf{RQ1:} Do Telegram Mini Apps comply with their stated privacy policies?
\item \textbf{RQ2:} How widely is Telegram's Standard Bot Privacy Policy used, and does providing a custom policy correspond to better compliance?
\item \textbf{RQ3:} At which stages of the Mini App lifecycle do privacy violations occur?
\item \textbf{RQ4:} What types of user data are predominantly shared with undeclared third parties?
\end{itemize}

To answer these RQs, we developed \toolName, a black-box dynamic analysis framework that requires no access to application code and takes as its only input the name of the bot associated with a Mini App. \toolName retrieves the applicable privacy policy, launches the Mini App from the Telegram client, intercepts its outbound traffic through a man-in-the-middle proxy, and separates traffic generated at page load from that produced during user-driven exploration. The contacted domains are then compared with the recipients declared in the applicable policy, either the custom notice exposed by the developer or, in its absence, Telegram's Standard Bot Privacy Policy.

We applied \toolName to 278 working Mini Apps sampled from tApps Center\footnote{\url{https://tapps.center/}}, a widely used community-driven catalog, with more than 4 million subscribers, that provides a large and publicly accessible entry point for discovering third-party applications in the Telegram ecosystem. We find that 59.4\% contact at least one third party that the applicable policy does not disclose, and that such contacts are seldom isolated: 63.0\% of violating apps reach two or more undeclared recipients, 2.78 on average and up to 14. The large majority of Mini Apps (78.8\%) do not expose a custom privacy notice and instead rely on the Standard Bot Privacy Policy; apps that provide a custom policy violate it at a statistically indistinguishable rate ($p = 0.55$). Violations are also front-loaded: 85.5\% of violating apps contact undisclosed third parties at page load, before any user action is possible. Across all 278 apps, not one presented a cookie banner, consent dialog, or other mechanism through which the user could accept, refuse, or configure the processing of personal data.

In summary, this paper makes the following contributions:
\begin{itemize}
\item We present the first empirical study of privacy-policy compliance in the Telegram Mini App ecosystem, covering 278 Mini Apps observed at runtime over five independent runs.
\item We propose \toolName, a dynamic analysis framework that detects undeclared third-party data flows in Mini Apps without access to application code, requiring only a bot name as input.
\item We derive from Telegram's Standard Bot Privacy Policy an explicit set of violation classes, making the default regime that governs 78.8\% of the analyzed ecosystem operational for automated analysis.
\item We demonstrate that the absence of consent mechanisms is widespread across the analyzed ecosystem and identify concrete platform-level countermeasures that Telegram could deploy.
\item We release our  Mini App scraper and dataset of 991 catalog entries to support replication and future work.\footnote{\url{https://github.com/Mobile-IoT-Security-Lab/TeleGapper}}
\end{itemize}

The remainder of this paper is organized as follows. Section~\ref{sec:back} introduces the Telegram Mini App architecture and the normative documents that govern it; Section~\ref{sec:rw} discusses related work; Section~\ref{sec:StandardBotPrivacyPolicy} derives our violation classes from the Standard Bot Privacy Policy; Sections~\ref{sec:methodology} and~\ref{sec:implementation} present the methodology and its implementation; Section~\ref{sec:exp} reports the experimental campaign and results; Section~\ref{sec:discussion} discusses the implications for the platform; Section~\ref{sec:Limitation} addresses threats to validity and limitations. Finally, Section~\ref{sec:conclusion} summarizes our contributions and outlines directions for future research.

\section{Background}
\label{sec:back}
This section summarizes the terminology and concepts required to understand the content of the rest of the paper.

\begin{figure}[h]
    \centering
\includegraphics[width=0.55\linewidth,height=0.6\textheight,keepaspectratio]{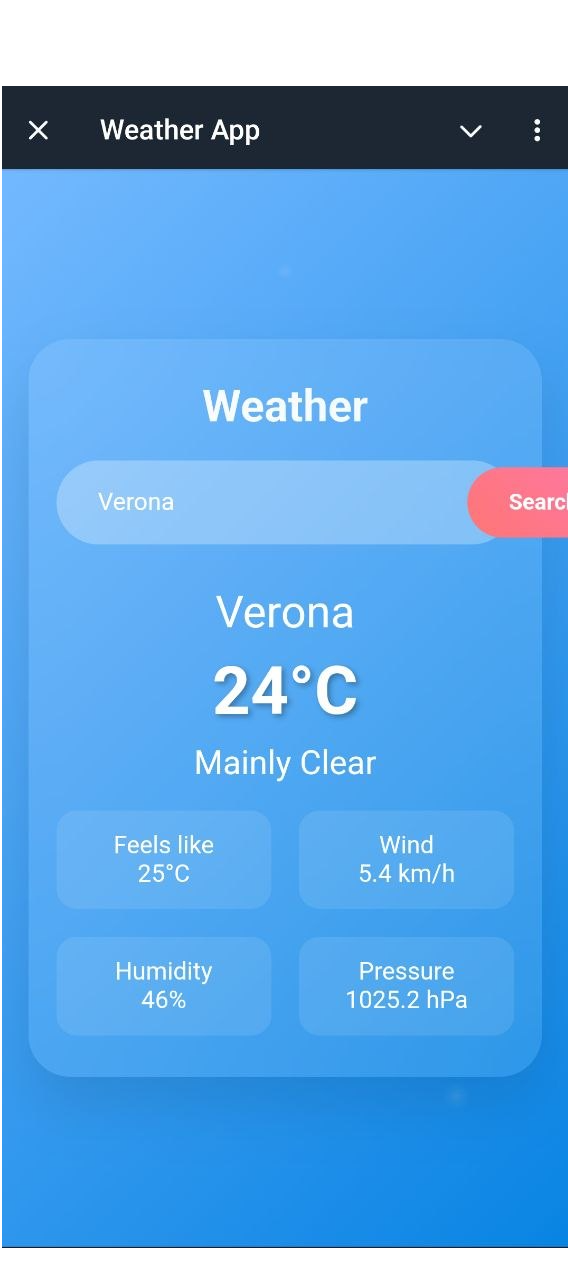}
    \caption{Telegram Mini App Front End Example}
    \label{fig:MiniApp}
\end{figure}

\paragraph{\textbf{The Telegram Mini App Architecture}}
\begin{figure}[h]
    \centering
    \includegraphics[width=\linewidth]{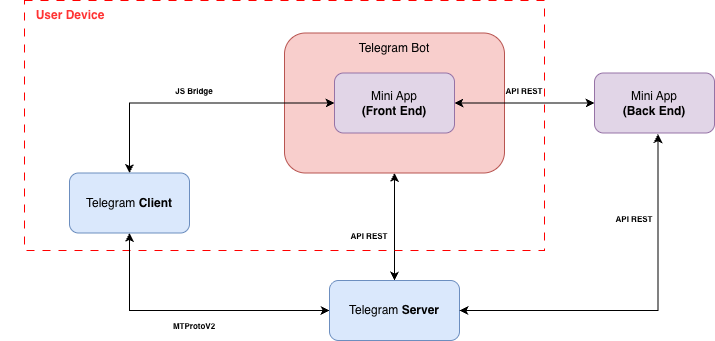}
    \caption{Telegram Mini App Execution Ecosystem}
    \label{fig:TelegramEcosystem}
\end{figure}
Compared to mini apps adhering to the W3C Mini App~\cite{MiniAppWhitePaper} standardization model, such as WeChat~\cite{WeChat}, Baidu~\cite{Baidu}, and Alipay~\cite{AliPay}, Telegram Mini Apps follow a fundamentally different architectural paradigm. Rather than being bundled into a single compressed package and executed in a controlled native-like environment, Mini Apps are full-fledged web applications hosted on third-party platforms (e.g., Vercel~\cite{Vercel}, Netlify~\cite{Netlify}). These applications are ultimately invoked via Telegram bots~\cite{TelegramBotApp}, which are automated software entities operating within the Telegram platform. They act as virtual chat partners that users interact with through text messages, predefined commands, inline queries, or interactive buttons, serving as entry points for launching the Mini Apps.

Due to the absence of a standardized packaging format or a mandatory application framework, the internal structure of a Telegram Mini App depends entirely on the developer’s technology stack. A Mini App may be implemented using plain HTML, CSS, and JavaScript, or using common frontend frameworks such as React~\cite{React} or Svelte~\cite{Svelte}, and is typically deployed on external hosting services. From an architectural perspective, it is useful to distinguish between a frontend component and a backend component. The frontend is the web application rendered inside the Telegram WebView and is responsible for the user interface, client-side logic, and interaction with the Telegram WebApp API, as shown in Figure~\ref{fig:MiniApp}. The backend, when present, is an external service under the developer’s control that exposes application APIs, executes server-side logic, validates Telegram initialization data, stores persistent data, and may interact with the Telegram Bot API on behalf of the service. For example, in a weather Mini App (Figure~\ref{fig:MiniApp}), the frontend may display the current temperature, weather conditions, and hourly or daily forecasts inside Telegram, while the backend handles API requests, retrieves meteorological data, manages user preferences, and stores saved locations.

\paragraph{\textbf{Telegram Mini App Execution Ecosystem}}
Telegram Mini Apps operate within a distributed execution ecosystem involving multiple components and communication paths. Figure~\ref{fig:TelegramEcosystem} illustrates the high-level architecture and key interactions.

\textit{1. Telegram Client.} The Telegram client, i.e., the application installed on the user's device, functions as a super-app, similarly to WeChat. It provides a runtime environment and a WebView container in which the Mini App is rendered and executed.

\textit{2. Telegram Bot.} The Telegram Bot~\cite{TelegramBot} serves as the interface between the user and the Mini App. It is responsible for triggering the app through specific commands or UI elements (e.g., inline buttons) and must be associated with the app to enable its launch~\cite{TelegramBot}.

\textit{3. Telegram Mini App.} The Mini App itself is a web application that executes within the Telegram WebView. It leverages standard web technologies (HTML, CSS, JavaScript) and communicates with the Telegram WebApp API to deliver interactive and context-aware experiences within the Telegram platform. As with conventional web applications, Mini Apps may also rely on a backend component to perform more complex server-side operations.

\textit{4. Telegram Server.}
The Telegram server exposes the Telegram Bot API, an HTTPS-based interface used by the bot backend to exchange data with the Telegram platform. Requests follow the format \url{https://api.telegram.org/bot<token>/METHOD_NAME}, where the bot token authenticates the bot and the method identifies the operation to execute, such as sending messages or retrieving updates. In the Mini App ecosystem, these requests are typically performed by the bot server or application backend, while the Mini App frontend running inside the Telegram WebView communicates primarily with its own backend and interacts with the Telegram client through the WebApp API, rather than directly invoking the Bot API.
\paragraph{\textbf{Mini App Market}}
Similar to  traditional Super Apps, such as WeChat, a user can retrieve a list of  mini apps by entering the keywords in a built-in  interface within the Telegram app. At a high level, this Telegram interface will return a list of the mini apps that match the keywords.
In addition to official discovery channels, the ecosystem contains alternative third-party marketplaces such as tApps Center~\cite{tApps} and FindiMini.app~\cite{FindMini}. These alternative platforms often serve as decentralized repositories that host a broader, unverified range of Mini Apps, providing users with access to experimental or community-driven projects that have not undergone Telegram's formal verification process. Unfortunately, the verification process is not publicly accessible, limiting transparency into how Mini Apps are reviewed and approved.
%\lv{questa parte è da modificare per quello che ci siamo detti in call}\lf{fixed}

\paragraph{\textbf{Telegram Policies}}
To govern both the use of the Telegram client and the operation of Mini Apps hosted on the platform, Telegram provides a set of interrelated normative documents: (i) \textit{the Telegram Privacy Policy}, (ii) \textit{the Terms of Service for Bots}, (iii) \textit{the Terms of Service for Mini Apps}, and (iv) \textit{the Standard Bot Privacy Policy}. Read together, these documents constitute the policy framework within which interactions among the user, the Telegram client, and Mini Apps and/or Bots occur. They also delimit the categories of user data that may be transmitted from the client to Mini Apps and their corresponding backend services, thereby defining the scope of data access, processing, and disclosure across the Telegram ecosystem.

\textit{1) Telegram Privacy Policy~\cite{TelegramPP}.}
Telegram does not provide a standalone privacy policy for Mini Apps; instead, it subsumes them within the broader bot ecosystem. The policy expressly states that the use of both Bot and Mini App features is governed, respectively, by the Terms of Service for Bots~\cite{TermsBot} and the Terms of Service for Mini Apps~\cite{TermsMiniApp}. In this framework, third-party developers are treated as independent entities from Telegram, and interactions with bots or related features may disclose public profile data, user messages, interface language, membership information in groups, and, when external links are accessed, the user’s IP address. Accordingly, the Privacy Policy establishes Telegram as the platform layer that mediates a limited set of data flows, while the operative privacy rules for Mini Apps are delegated to the dedicated terms governing those services.

\textit{2) Terms of Service for Bots~\cite{TermsBot}.}
The Terms of Service for Bots explicitly place Mini Apps within the same third-party ecosystem as bots. They further state that access to a Mini App is additionally governed by the Terms of Service for Mini Apps, while Telegram disclaims liability for the availability, operation, and loss of any data, assets, or functionality delivered through either Bots or Mini Apps, which remain under the responsibility of the respective third-party service provider.

\textit{3) Terms of Service for Mini Apps~\cite{TermsMiniApp}.}
The Terms of Service for Mini Apps further specify the privacy and liability regime applicable to Mini Apps. Telegram states that Mini Apps are third-party services operated by independent Service Providers, who are solely responsible for their operation, maintenance, content, and availability. From a data-processing perspective, a Mini App automatically acquires the user’s IP address and may receive basic account metadata, including the Telegram user ID, public name, username, profile picture, client language tag, premium subscription status, and selected in-app theme parameters. If launched from a private chat, it may also receive basic information about the chat partner; if opened from channels or group chats, it may additionally obtain the ID, type, title, username, and photo of the corresponding chat. Telegram further specifies that such data are shared with the relevant Service Provider and that, once transmitted, Telegram no longer controls the subsequent exchange of information between the user and the provider. Payments are also handled by third-party payment providers, while Telegram disclaims responsibility for disputes, refunds, service quality, availability, and any losses arising from transactions conducted through Mini Apps.

\textit{4) Standard Bot Privacy Policy~\cite{MiniAppPP}.}
The Standard Bot Privacy Policy functions as a default privacy notice for third-party bots and Mini Apps hosted on the Telegram platform and applies in the absence of a separate privacy policy disclosed by the Mini App developer. It clarifies that such services are independent third-party applications that are neither maintained nor endorsed by Telegram, and that the policy governs exclusively the relationship between the Developer and the User, without replacing the Telegram Privacy Policy.

\section{Related Works}

\label{sec:rw}

Prior work has extensively investigated privacy and security issues in mini app ecosystems, particularly through static and taint-based analyses aimed at detecting sensitive data flows and inconsistencies between application behavior and declared privacy policies. However, most of this literature focuses on WeChat and other Chinese mini app ecosystems, while Telegram Mini Apps remain comparatively underexplored.

Several studies target privacy leakage and policy compliance. TaintMini~\cite{wang2023taintmini} constructs a universal data-flow graph to identify tainted flows within and across mini apps. Wang et al.~\cite{wang2024you} analyze AST nodes to extract data types and operations and use a matching model to assess consistency between application code and privacy policies. MiniTracker~\cite{li2023minitracker} instead builds an assignment flow graph (AFG) that integrates JavaScript and HTML data flows and applies property-based taint analysis to track sensitive information at finer granularity. These approaches demonstrate the effectiveness of code-level analysis for detecting privacy-relevant behavior, but fundamentally rely on the availability of inspectable mini app artifacts.

A parallel line of work investigates security vulnerabilities in mini app ecosystems~\cite{liu2020industry,wang2022characterizing,yang2022cross}. Wang et al.~\cite{wang2022characterizing} characterize common bugs in WeChat mini apps and develop automated detection techniques, while Yang et al.~\cite{yang2022cross} identify cross-mini app request forgery attacks. Other studies show that vulnerabilities can originate from the host platform itself~\cite{lu2020demystifying,wang2023one,zhang2022identity,wang2023uncovering}. For example, Zhang et al.~\cite{zhang2022identity} identify authentication weaknesses that may lead to identity-confusion attacks and unauthorized access to privileged APIs, whereas Wang et al.~\cite{wang2023uncovering} uncover undocumented and insufficiently protected host APIs that may allow mini apps to bypass platform restrictions.

More recently, MiniEval~\cite{xiang2026minieval} extends privacy-oriented analysis by combining automated compliance-violation detection with quantitative privacy-risk assessment. Together, these works provide increasingly sophisticated techniques for analyzing privacy and security in conventional mini app ecosystems. Their applicability to Telegram, however, is limited by a fundamental architectural difference: Telegram Mini Apps are Web applications dynamically loaded from developer-controlled URLs rather than packaged artifacts that can be statically inspected before execution.

Research specifically targeting Telegram Mini Apps remains limited. Mohammadi et al.~\cite{mohammadi2025security} analyze security vulnerabilities and exploits in Telegram Mini Apps and propose a dedicated mitigation framework that focuses on architectural weaknesses such as unproxied communication, insecure HTTP configurations, WebView-related attacks, and general privacy leakage. Their work, however, does not systematically compare runtime third-party communications with the privacy disclosures presented to users, nor does it investigate when potentially undisclosed communications occur with respect to user interaction and consent.

Our work addresses this complementary gap. Rather than inspecting application code or focusing on exploitable vulnerabilities, we analyze Telegram Mini Apps as black boxes at runtime. To the best of our knowledge, this is the first empirical study to systematically assess whether their observed third-party communications are consistent with the applicable privacy policies and whether such communications occur before users are given an opportunity to provide consent.

\section{Standard Bot Privacy Policy Analysis}

\label{sec:StandardBotPrivacyPolicy}
%\lv{Rivista la sezione per allineare meglio il testo alla Standard Bot Privacy Policy e a ciò che TeleGapper può effettivamente osservare. In particolare, ho chiarito la definizione di \textit{Third-Party Service}, precisato i vincoli su necessità e condivisione dei dati, e reso le classi di violazione più aderenti agli indicatori realmente ricavabili dal traffico di rete, evitando inferenze troppo forti su retention, purpose deviation e ruolo di data controller.} \lf{OK}
The \textit{Standard Bot Privacy Policy} defines the default data-handling regime for Telegram third-party bots and Mini Apps when no developer-specific privacy notice is provided. Its role is not merely descriptive: it establishes a baseline of permitted data collection, processing, and disclosure for services relying on the default policy.

The policy defines a \textit{Third-Party Service} as the bot or Mini App operated by the developer~\cite{MiniAppPP}. From a data-processing perspective, such services may access the limited set of user information exposed through Telegram and process it only insofar as necessary for their designated features. This introduces an explicit necessity constraint: the mere availability of user information does not, by itself, justify its collection or processing. Identifiers, profile metadata, and initialization data should therefore be processed only when relevant to the functionality provided by the service.

The policy further constrains downstream disclosure. Private user information must not be transferred or made accessible to third parties unless such sharing is explicitly authorized by the user or otherwise permitted by the policy or applicable law~\cite{MiniAppPP}. Consequently, outbound communications with external analytics, advertising, tracking, or auxiliary services become policy-relevant when they involve user-related information and no corresponding authorization or disclosure can be established.

The policy also imposes purpose-limitation and data-minimization requirements. Developers may not use user data outside the scope of the Third-Party Service unless such use is clearly stated and explicitly agreed to by the user, and data collection must remain limited to what is necessary to provide or enhance the service functionality~\cite{MiniAppPP}. The developer is further responsible for handling, transferring, and storing user information in accordance with applicable law. We therefore treat processing that is unrelated to the declared service functionality, or sharing with undeclared external recipients, as potentially inconsistent with the default policy.

Finally, the policy grants users rights over personal information collected and stored by the Third-Party Service, including access, deletion, restriction, objection, and withdrawal of previously given consent~\cite{MiniAppPP}. Although these requirements cannot be fully evaluated through network observation alone, they reinforce the policy's broader expectation that processing should remain transparent, limited, and accountable.

Table~\ref{tab:policy-violations} summarizes the policy constraints that can be operationalized through our dynamic analysis. The first column identifies the policy-level violation class, the second reports the corresponding observable network-level indicator, and the third explains how that indicator relates to the requirements of the Standard Bot Privacy Policy. Importantly, we distinguish direct evidence of a policy mismatch from indicators that only motivate further investigation.

For example, \textit{Undisclosed third-party communication} captures requests to external domains that are not identified by the applicable privacy policy. When such communications involve user-related information, they provide direct evidence that data are disclosed to recipients not reflected in the policy. Similarly, \textit{Pre-interaction third-party communication} captures external requests generated at page load, before any meaningful user interaction. While timing alone does not establish a violation, it is particularly relevant when the transfer would require user authorization, because no application-level interaction has yet occurred through which such authorization could have been obtained.

\begin{table*}[t]
\centering
\small
\setlength{\tabcolsep}{5pt}
\renewcommand{\arraystretch}{1.15}
\caption{Violation classes derived from the Standard Bot Privacy Policy and their observable network-level indicators.}
\label{tab:policy-violations}
\begin{tabularx}{\textwidth}{>{\raggedright\arraybackslash}X
                            >{\raggedright\arraybackslash}X
                            >{\raggedright\arraybackslash}X}
\toprule
\textbf{Violation class} & \textbf{Network-level indicator} & \textbf{Interpretation under policy constraints} \\
\midrule
\textbf{Undisclosed third-party communication} &
Requests to external domains not identified in the applicable privacy policy &
Indicates communication with recipients that are not disclosed to the user and may therefore fall outside the declared processing scope. \\

\addlinespace
\textbf{Third-party transmission of user data} &
User identifiers or profile attributes transmitted to an external domain &
Indicates disclosure of user-related information to a third party, which requires an applicable policy basis or user authorization. \\

\addlinespace
\textbf{Potential over-collection} &
Transmission of Telegram-provided attributes not evidently required by the observed service functionality &
May indicate processing beyond what is necessary for the designated features of the Mini App. \\

\addlinespace
\textbf{Pre-interaction third-party communication} &
External requests generated at page load before meaningful user interaction &
Identifies data flows occurring before the user can interact with the Mini App or provide application-level authorization. \\

\addlinespace
\textbf{Tracking or advertising communication} &
Requests to domains classified as analytics, advertising, or tracking services &
Identifies processing that may fall outside the core service functionality and therefore requires explicit disclosure or authorization when user data are involved. \\
\bottomrule
\end{tabularx}
\end{table*}

%\section{TeleGapper: Design and Analysis Workflow}
\section{TeleGapper: Design and Analysis Workflow}
%\lv{Riorganizzata la sezione per separare meglio il workflow concettuale dai dettagli implementativi. Ho mantenuto qui solo logica dell’analisi, distinzione tra initialization e interaction, confronto con la privacy policy, repeated runs, review manuale e consent inspection, spostando tempi, parametri ed euristiche tecniche nella sezione di implementation.}\lf{OK}
\label{sec:methodology}

\begin{figure*}[!ht]
    \centering
    \includegraphics[width=\textwidth]{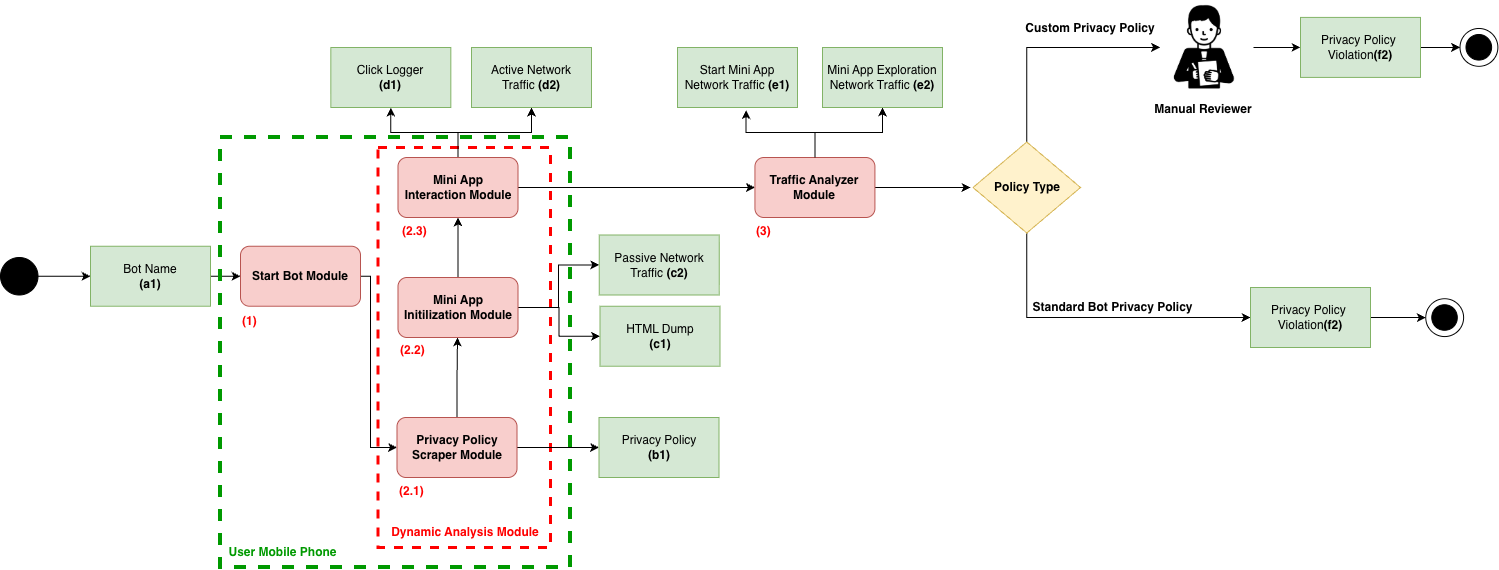}
    \caption{Workflow of \toolName{} for detecting privacy-policy violations and analyzing the timing of user-data transmissions.}
    \label{fig:Methodology}
\end{figure*}

\subsection{Framework Overview}

We present \toolName, a black-box dynamic analysis framework for detecting privacy-policy violations in Telegram Mini Apps. Given only the name of the bot associated with a Mini App, \toolName analyzes the application at runtime by combining network-traffic inspection with the applicable privacy policy, without requiring access to its source code.

As illustrated in Figure~\ref{fig:Methodology}, the workflow consists of three main components: (i) the \textit{Start Bot Module}, (ii) the \textit{Dynamic Analysis Module}, and (iii) the \textit{Traffic Analyzer Module}. At a high level, \toolName identifies the target bot, retrieves the privacy policy associated with its Mini App, launches the application from the Telegram client, and performs black-box exploration while intercepting its network traffic. The collected traffic is then analyzed to identify user-data transmissions and compared with the applicable privacy policy. Notably, the workflow distinguishes traffic generated during Mini App initialization from traffic triggered by subsequent user interaction, allowing us to determine when privacy-relevant communications occur.

\subsection{Bot Discovery and Privacy-Policy Retrieval}

\paragraph{Start Bot Module}
Given a bot name or a list of bot names \textbf{(a1)}, the \textit{Start Bot Module \textbf{(1)}} automatically opens the Telegram client, searches for the target bot, and opens the corresponding chat.

\paragraph{Privacy Policy Scraper Module}
Once the bot chat is opened, the \textit{Privacy Policy Scraper Module \textbf{(2.1)}} accesses the bot profile, navigates to the ``More options'' section, and retrieves the privacy policy exposed through the ``Privacy Policy'' entry. The retrieved policy is stored as artifact \textbf{(b1)} and is later used by the Traffic Analyzer Module to determine which disclosure regime applies to the Mini App.

\subsection{Dynamic Mini App Analysis}

The \textit{Dynamic Analysis Module} separates Mini App execution into two stages: initialization and user-driven exploration. This separation is central to our analysis because it allows us to distinguish network activity generated automatically at page load from activity triggered after the user starts interacting with the application.

\paragraph{Mini App Initialization Module}
The \textit{Mini App Initialization Module \textbf{(2.2)}} launches the Mini App from the bot profile and records all network traffic generated during the loading phase \textbf{(c2)}, without performing any user interaction. After allowing the application to complete its initial loading phase, the module extracts the HTML content of the rendered page \textbf{(c1)}. The HTML snapshot supports subsequent exploration and is also used to inspect the initial application state.

Separating initialization from later interaction is important because transmissions observed during this phase occur before the user can meaningfully interact with the Mini App. Such traffic, therefore, provides evidence about data flows that arise automatically at launch and can be analyzed independently from user-triggered behavior.

\paragraph{Mini App Interaction Module}
Following initialization, the \textit{Mini App Interaction Module \textbf{(2.3)}} explores the Mini App using the previously extracted HTML structure \textbf{(c1)} within a bounded exploration budget. Because Mini Apps may rely on standard DOM elements, \texttt{canvas}-based rendering, or a combination of both, \toolName adopts three exploration modes:

\begin{itemize}
    \item \textit{Classic DOM Analysis}: the module identifies clickable elements on the current page and activates them in randomized order, returning to the previous state after each interaction when possible.
    \item \textit{Canvas-Only Analysis}: if no clickable DOM elements are available and a visible canvas is detected, the module partitions the visible canvas into a grid and performs randomized interactions across its cells.
    \item \textit{Hybrid Canvas and DOM Analysis}: when both clickable DOM elements and a visible canvas are present, the exploration budget is divided between Classic DOM Analysis and Canvas-Only Analysis.
\end{itemize}

During exploration, the module records all performed interactions in JSON format \textbf{(d1)} to support reproducibility and captures the outbound network traffic generated by user-driven interaction \textbf{(d2)}. In contrast to the initialization phase, traffic observed here may depend on actions performed after the Mini App has become interactive.

\subsection{Traffic and Policy Analysis}

Once dynamic exploration completes, the \textit{Traffic Analyzer Module \textbf{(3)}} processes the traffic collected during initialization \textbf{(c2)} and user-driven exploration \textbf{(d2)} and produces two structured reports.

The first report \textbf{(e1)} describes traffic observed during Mini App initialization. It identifies user-related data transmitted to external services, the corresponding destination endpoints, and whether such transmissions occur before the user can meaningfully interact with the Mini App. The second report \textbf{(e2)} applies the same analysis to traffic generated during user-driven exploration.

For each observed transmission, the reports record the detected data type, value, and destination endpoint. This representation allows us to determine which user-related data are transmitted, to which endpoints, and at which stage of the Mini App lifecycle the communication occurs.

The resulting network reports are then evaluated against the applicable privacy policy \textbf{(b1)}. When the Mini App relies on Telegram's Standard Bot Privacy Policy, the observed flows are assessed according to the policy-derived violation classes defined in Section~\ref{sec:StandardBotPrivacyPolicy}. When a custom privacy policy is exposed, the reports are reviewed against that policy to determine whether the observed third-party recipients and data-sharing practices are disclosed.

We define a \emph{privacy-policy violation} as the transmission of user data to a third-party domain that is not disclosed by the privacy policy applicable to the Mini App, namely the custom policy exposed by the developer or, in its absence, Telegram's Standard Bot Privacy Policy.

\subsection{Repeated Observation and Review Protocol}

\textbf{Repeated Observation.}
The exploration strategy is randomized, and a single execution may therefore cover only a subset of a Mini App's reachable states and generated traffic. We consequently repeat the complete workflow (i.e., launch, initialization capture, interaction, and traffic capture) $5$ times for each Mini App, restarting the Telegram client and clearing the WebView state between runs. Repetitions are used to increase behavioral coverage rather than to filter observations. We therefore take the \emph{union} of the traffic observed across runs: a third-party contact is retained for a given Mini App and execution stage whenever it is observed in at least one run.

\textbf{Manual Review of Custom Policies.}
When a Mini App exposes a custom privacy policy, the reviewer follows an annotation protocol similar to~\cite{ferrari2026policygapperautomateddetectioninconsistencies}. The reviewer (i) reads the complete policy and marks statements describing data-collection or data-sharing practices and (ii) compares those statements with the third-party requests reported in \textbf{e1} and \textbf{e2}. A recipient is considered disclosed whenever the policy can plausibly be read as covering it; otherwise, the corresponding flow is recorded as a privacy-policy violation \textbf{(f2)}.

\textbf{Consent Artifact Inspection.}
For each Mini App, we independently inspect the initial interface rendered at launch for observable consent artifacts, including cookie banners, consent dialogs, privacy notices requiring acknowledgment, or controls through which the user can accept, refuse, or configure data processing. This inspection is performed independently of the randomized exploration so that the interaction strategy cannot overlook an initially visible consent element. This step does not by itself determine whether a privacy-policy violation occurred; instead, it establishes whether traffic observed during initialization \textbf{(c2)} is preceded by an observable opportunity for the user to express a preference.

%\section{Implementation}
\section{Implementation Details}
\label{sec:implementation}
%\lv{Rinominata e ristrutturata la sezione come Implementation Details, concentrandola sulle scelte ingegneristiche: Android/Appium, proxy Burp, device setup, tempi di analisi, strategia canvas, parsing del traffico, regex, normalizzazione, filtri EXCLUDED\_HOSTS e formato degli output. Ho anche reso più chiaro il ruolo separato dello scraper rispetto al framework principale.}\lf{OK}

This section describes the implementation of \toolName and the main engineering choices underlying our Android-based analysis pipeline.

\subsection{Dataset Scraper}

To build the dataset used in the experimental campaign (Section~\ref{sec:exp}), we developed a dedicated scraper for tApps Center\footnote{\url{https://github.com/Mobile-IoT-Security-Lab/TeleGapper}}. The scraper is implemented in Python on top of Selenium\footnote{\url{https://www.selenium.dev/}}, which drives a headless browser and allows us to collect catalogue entries rendered client-side, for which a plain HTTP-based crawler would retrieve no complete listing.

The scraper enumerates the catalogue published by the portal and, for each entry, extracts the associated bot name together with the app title and category. The bot name is the only input required by the \textit{Start Bot Module} (\textbf{a1}), making the scraper the entry point of the data-collection pipeline.

We keep this component separate from the analysis framework because it plays no role in the detection of privacy-policy violations and operates on the tApps web portal rather than on the mobile client, relying on a different technology stack. Releasing it as a standalone artifact also allows it to be reused independently to obtain an updated snapshot of the applications listed by the portal.

We selected tApps Center as the source of our sampling frame because it provides a large, publicly accessible, and structured catalogue for discovering third-party applications in the Telegram ecosystem. Its client-rendered listings expose bot names and associated metadata and can be collected automatically at scale using a headless browser. The catalogue also organizes Mini Apps into functional categories, allowing the sampled population to span different service domains rather than a single application type.

\subsection{Mobile Automation and Exploration}

%\paragraph{Start Bot and Dynamic Analysis Modules.}
The \textit{Start Bot Module} and \textit{Dynamic Analysis Module} are implemented using Appium\footnote{\url{https://appium.io/docs/en/latest/}}, which enables automated interaction with Telegram and its embedded WebView-based Mini Apps. Appium was selected because it can drive hybrid mobile applications and expose both native and web-based UI elements. Given a bot name, the framework launches Telegram, locates the bot, opens the corresponding Mini App, retrieves the Privacy Policy, and performs interaction-driven exploration.

The Mini App initialization phase lasts 20 seconds before the HTML snapshot is collected. This interval was selected to provide the application sufficient time to complete its initial loading behavior while keeping the analysis scalable.

The subsequent interaction phase uses a 240-second exploration budget. This duration was selected empirically as a trade-off between behavioral coverage and analysis cost. For Canvas-Only Analysis, the visible canvas is partitioned into a grid of 8 rows and 4 columns. The module repeatedly selects a random cell and clicks a random position inside it, reducing spatial clustering while maintaining broad screen coverage. In Hybrid Canvas and DOM Analysis, the exploration budget is split evenly between DOM-based and canvas-based interaction.

\subsection{Network Traffic Interception}

%\paragraph{Proxy.}
To intercept and inspect network traffic generated by Mini Apps, we use Burp Suite\footnote{\url{https://portswigger.net/burp}} as a man-in-the-middle (MITM) proxy. The analysis is performed on a rooted Google Pixel 9 configured with Magisk, allowing system-wide certificate installation and HTTPS traffic interception. The device routes its traffic through a workstation running Burp Suite, on which a custom CA certificate is installed to enable TLS decryption.

Network traces are exported from Burp Proxy in raw HTTP format and stored as plain-text logs. This design avoids reliance on Burp-specific APIs and allows the subsequent analysis stage to operate on portable artifacts.

\subsection{Traffic Analysis Pipeline}

%\paragraph{Traffic Analysis Module.}
The \textit{Traffic Analysis Module} is implemented as a post-processing pipeline that consumes the raw HTTP/HTTPS traces exported by the proxy and converts them into structured JSON artifacts. Rather than relying on proxy-specific APIs, the module parses the plain-text logs and reconstructs individual requests by extracting the HTTP method, URL, host, headers, body, timestamp, and destination IP address.

To separate first-party Mini App traffic from external communications, the module infers the Mini App domain by combining multiple traffic signals, including the \texttt{Origin} and \texttt{Referer} headers and Telegram-specific initialization parameters such as \texttt{tgWebAppData}, \texttt{initData}, \texttt{query\_id}, and \texttt{auth\_date}. Requests matching the inferred Mini App domain are excluded from third-party inspection, while the remaining requests are treated as candidates for further analysis.

The pipeline then applies a set of regular-expression extractors to third-party requests to identify potentially sensitive information. For example, the following pattern detects Telegram Mini App initialization data:

\begin{verbatim}
r"(?<![a-zA-Z0-9_])(?:tgWebAppData|initData|
init_data|initdata)(?:=|[\"']?\s*:\s*[\"']?)
([^\"'\s<]{8,})"
\end{verbatim}

The expression captures initialization parameters across common serialization formats, including URL-encoded queries, JSON payloads, and loosely structured text. The negative lookbehind prevents matches that are part of longer identifiers, while the non-capturing group covers multiple naming variations. The separator pattern supports both URL-style key-value pairs and JSON-style representations, and the final capturing group extracts the corresponding value while requiring a minimum length to reduce trivial matches.

For example, a request may contain an initialization fragment of the form:

\begin{verbatim}
tgWebAppData=user={"id":XXXXX,"first_name":
"XXXXX","last_name":"XXXXX","username":"XXXXX",
"language_code":"en"}&auth_date=XXXXXX...
\end{verbatim}

The extractor recovers the initialization payload, including user attributes and authentication metadata, which is then forwarded to subsequent decoding and privacy-assessment stages.

The extractors cover Telegram identifiers, profile metadata, authentication tokens, API keys, cookies, device and browser metadata, visited-page URLs, location coordinates, analytics identifiers, and TON wallet addresses. Before matching, request contents are normalized through recursive decoding of URL-encoded values, extraction of query parameters and fragments, parsing of form-encoded payloads, flattening of JSON structures, and selective decoding of Base64-encoded content. This normalization improves detection coverage for values embedded in complex or nested payloads.

\subsection{Background-Traffic Filtering}

To reduce background noise, the implementation maintains an \texttt{EXCLUDED\_\-HOSTS} set containing infrastructure and runtime-generated domains that are not attributable to the analyzed Mini App. The set was constructed by manually inspecting traffic from 100 Mini Apps randomly sampled from tApps Center and retaining hosts that consistently appeared across unrelated applications and could therefore be attributed to the underlying Android or platform environment rather than to app-specific behavior.

Requests targeting these hosts are filtered before third-party inspection. The excluded set includes, among others, Android and Google infrastructure used for connectivity checks, device verification, system libraries, and public assets. Examples include \texttt{android.googleapis.com}, \texttt{play.googleapis.com}, \texttt{fonts.googleapis.com}, \texttt{connectivitycheck.gstatic.com}, and \texttt{on\-device\-safety-pa.googleapis.com}.

This filtering step reflects the scope of our analysis: we aim to characterize communications attributable to Mini App behavior rather than operating-system background traffic. We therefore exclude hosts that consistently arise independently of the analyzed application. This choice is revisited as a potential source of false negatives in Section~\ref{sec:Limitation}.

\subsection{Generated Artifacts}

For each analyzed Mini App, the Traffic Analysis Module outputs two JSON reports, \textbf{e1} and \textbf{e2}, corresponding respectively to initialization traffic and interaction-generated traffic. Each report includes the inferred Mini App domain and the detected user-related data transmitted to external destinations. These reports are subsequently evaluated against the applicable privacy policy as described in Section~\ref{sec:methodology}.

\section{Experimental Campaign and Analysis Results}
\label{sec:exp}
The experimental campaign is supported by one dataset, $D_{tApps}$, which comprises all Mini Apps available on tApps Center\footnote{\url{https://tapps.center/}}.
$D_{tApps}$ consists of 991 catalogue entries, collected in the first half of June 2026 with the Mini App Scraper described in Section~\ref{sec:implementation}.
We stress that the scraper collects the listing published by the portal, i.e.\ the bot name and the associated metadata, and not the application itself: whether the corresponding Mini App is still deployed and reachable can only be established at analysis time, by opening it from the Telegram client.

To answer the research questions introduced in Section~\ref{sec:intro} (RQ1--RQ4), we applied our methodology to a random sample of $D_{tApps}$. Manual inspection of such a sample is commonly employed in empirical software engineering studies~\cite{ferrari2026evaluatingllmsobfuscationdetection,alecci2025damflow,alecci2026taskflow}.% \lv{troppe citazioni, max 3 le più rilevanti}\lf{fixed}.
Using a confidence level of 95\% and a margin of error of 5\% over the 991 entries of $D_{tApps}$, we set a target of 278 Mini Apps to be manually inspected.
We then drew entries from $D_{tApps}$ uniformly at random, without replacement, across all Mini App categories, and submitted each of them to the pipeline described in Section~\ref{sec:methodology}, discarding those for which no runtime behaviour could be observed and drawing a replacement, until the target was met.
An entry was discarded according to the following criteria, each verified manually on the device after the automated attempt had failed, so as to rule out transient network or timing errors:

\begin{itemize}
    \item \emph{Deleted bot} (27 entries). The bot name published in the catalogue yields no result when searched in the Telegram client, as the developer has deleted the bot after its listing was approved. The tApps Center is evidently not synchronised with the actual availability of the bots it advertises, an observation we return to in Section~\ref{sec:discussion}.
    \item \emph{Unreachable frontend} (124 entries). The bot exists and the Mini App can be launched, but the WebView fails to render the application, as the domain serving the frontend no longer resolves, has expired, or returns an HTTP error.
    \item \emph{No profile-level entry point} (13 entries). The bot exists and is reachable, but its profile does not expose the \textit{Open App} shortcut on which our \emph{Start Bot Module} relies, so the Mini App cannot be launched by our automation. Unlike the two categories above, these entries are not evidence of an inactive Mini App, but of a coverage limitation of our tool, which we discuss in Section~\ref{sec:extValidity}.
\end{itemize}

Reaching the target of 278 analyzable Mini Apps, therefore, required drawing 442 entries, 164 of which (37.1\%) were discarded. The results reported in the remainder of this section are computed over the 278 Mini Apps that were successfully launched and observed at runtime.
The experimental campaign was carried out on a 12th Gen Intel(R) Core(TM) i9-12900KS
(3.40 GHz) server with 128 GB RAM, and on a rooted
Google Pixel 9 (Android 16) with Telegram 12.7.3.
The use of a rooted device does not affect the validity of the analysis: Telegram's clients are open source and expose developer/debug options, which explicitly allow inspection and debugging of the application at user level.
The device is rooted only to install our CA certificate in the system trust store, a precondition for decrypting HTTPS traffic on Android~7 and later (Section~\ref{sec:implementation}). Root is therefore used to observe traffic, not to alter execution: the Telegram client is neither patched nor instrumented, and Mini Apps run in the same WebView they would on a stock device.
The results of this analysis are presented in the following sections.

\subsection{RQ1: Do Telegram Mini Apps actually comply with their stated privacy policies?}
\label{sec:rq1}
For each working app, we checked whether at least one violation, as defined in Section~\ref{sec:methodology}, occurred at either of the two monitored stages, i.e.\ at opening or during use.  As shown in Table~\ref{tab:violation_distribution},   more than half of the analyzed apps (59.4\%) exhibit at least one violation with respect to what is declared in their own privacy policies.

\begin{table}[h]
\centering
\caption{Privacy policy violations among the 278 working Mini Apps. An app counts as violating if it contacts an undeclared third party at opening, during use, or both.}
\label{tab:violation_distribution}
\resizebox{\columnwidth}{!}{%
\begin{tabular}{lcc}
\toprule
\textbf{Category} & \textbf{N} & \textbf{\%} \\
\midrule
With at least one violation (at opening or during use) & 165 & 59.4\% \\
No violation detected & 113 & 40.6\% \\
\bottomrule
\end{tabular}%
}
\end{table}

We manually classified every third-party domain observed in the violating requests. Each domain was first attributed to the service and the provider operating it, which we then looked up in AppBrain~\cite{AppBrainADV}, consulted in June 2026, a publicly available catalogue that indexes third-party libraries and SDKs and assigns each of them to a category, such as ad network, analytics, social, or development tool. AppBrain
was adopted as the primary reference because it provides an externally maintained categorisation, which makes the classification reproducible and independent of our own judgement. When a service had no entry in the catalogue, which was typically the case for providers specific to the Telegram ecosystem, we performed a manual web search for the corresponding domain and determined the service category from the provider's official website or documentation, recording the purpose declared therein.
%we resorted to a manual web search on the domain itself and determined the nature of the service from the official website and the developer documentation of its provider, recording the purpose declared therein. 

This classification reveals that violations are concentrated in a small, clearly identifiable set of tracking and advertising services. Table~\ref{tab:third_parties} reports the five most frequently contacted third parties at app opening and during use, respectively.

The pattern that emerges from Table~\ref{tab:third_parties} concerns who is
contacted, what kind of service they are, and when the contact occurs. The
recipients are few and recurring: only eight distinct services occupy the ten
top-five positions, and two of them, Google Tag Manager and Yandex Metrica,
appear at both stages, always within the top three. All eight are analytics,
profiling or ad-serving services; none is a generic technical component or a
diagnostic instrument such as a crash reporter. Adsgram, Monetag and Rtmark are
advertising networks specific to or commonly used in the Telegram Mini App ecosystem, whereas Google
Tag Manager, Google Marketing, Yandex Metrica, Meta Analytics and TikTok
Analytics belong to general-purpose tracking and ad-serving infrastructures.
The two stages, however, are dominated by different actors. 
The Telegram-specific advertising networks hold three of the five top positions at opening and none during use, where the general-purpose analytics platforms take over, led by Yandex Metrica and Google Tag Manager. A large share of this traffic, moreover, precedes any user interaction: Google Tag Manager is contacted at launch in 46 of the 165 violating Mini~Apps (27.9\%) and Adsgram in 40 (24.2\%).
Thus, for roughly one quarter of violating apps, communication with an undeclared tracking or advertising service begins before any user interaction.
%In roughly a quarter of the violating apps, undeclared tracking thus begins before the user has done anything at all.

\begin{table}[h]
\centering
\caption{Five most frequently contacted third parties at opening and during
use, across the 165 violating Mini~Apps, with the number of Mini~Apps in which
each was observed. Counts are per stage: the same Mini~App may contribute to
both columns.}
\label{tab:third_parties}
\resizebox{\columnwidth}{!}{%
\begin{tabular}{lc|lc}
\toprule
\textbf{Third party (opening)} & \textbf{Occ.} & \textbf{Third party (runtime)} & \textbf{Occ.} \\
\midrule
Google Tag Manager & 46 & Yandex Metrica     & 33 \\
Adsgram            & 40 & Google Tag Manager & 31\\
Yandex Metrica     & 17 & Meta Analytics            & 19\\
Rtmark            & 16 &  TikTok Analytics    & 13 \\
Monetag &  14 &  Google Marketing & 11 \\
\bottomrule
\end{tabular}%
}
\end{table}

%Their systematic recurrence across unrelated applications suggests a structural phenomenon rather than a set of isolated or incidental errors. The typical violation observed does not consist of a marginal technical defect, but of a recurring pattern of data sharing with third-party advertising brokers, undeclared in the privacy policy. This behaviour is relevant not only from a technical standpoint but also from a regulatory one: communicating personal data to third parties not identified in the privacy notice constitutes a direct violation of the transparency obligations set out in Articles~13 and~14~\cite{art13,art14} of the GDPR.
Their systematic recurrence across unrelated applications suggests that these violations are not limited to isolated implementation mistakes, but reflect a recurring pattern of undeclared third-party communication. In many cases, the observed violations involve communications with advertising or tracking services that are not disclosed in the applicable privacy policy. This behaviour is relevant not only from a technical standpoint but also from a regulatory one: Articles~13 and~14 GDPR~\cite{art13,art14} require users to be informed about the recipients, or categories of recipients, of their personal data. Accordingly, transmitting personal data to recipients that are not covered by the applicable privacy notice may be inconsistent with these transparency obligations. %\lv{rivisto paragrafo da rileggere}

Table~\ref{tab:tp-per-app} reports, for each violating Mini App, the number of distinct undeclared third-party domains contacted at either monitored stage. In 61 cases (37.0\%) the violation involves a single recipient, while the remaining 104 apps (63.0\%) contact two or more undeclared third parties; 30 of them (18.2\%) contact five or more, and the maximum observed in our sample is 14. On average, a violating Mini App discloses data to 2.78 undeclared third parties (median 2). The distribution is therefore markedly skewed: most violations are not isolated communications with a single forgotten endpoint, but involve several distinct recipients at once, with a long tail of applications contacting multiple tracking or advertising services.%leaks towards a single forgotten endpoint, but involve several distinct recipients at once, with a long tail of applications embedding an entire stack of trackers.

\begin{tcolorbox}[colback=gray!5,colframe=blue!40, boxrule=0.3mm, arc=0mm, left=1mm, right=1mm, top=1mm, bottom=1mm]
\textbf{RQ1: Do Telegram Mini Apps actually comply with their stated privacy policies?}

More than half of the working Mini Apps analyzed (59.4\%, 165/278) were observed
communicating with at least one third party not disclosed in their privacy policy.
Violations are also seldom confined to a single recipient: 63.0\% of the violating
apps contact two or more distinct undeclared third parties, with an average of 2.78
per app (median 2, up to 14). Rather than isolated technical errors, these
violations concentrate on a small, recognizable set of tracking and advertising
services (Google Tag Manager, Adsgram, Yandex Metrica, Meta Analytics, TikTok
Analytics), suggesting a recurring pattern of undeclared data sharing.
\end{tcolorbox}
\begin{table}[t]
\centering
\caption{Number of undeclared third parties per violating Mini App.
Counts refer to distinct third-party domains contacted at either
monitored stage (opening or during use), aggregated per app.
Percentages are computed over the 165 violating Mini Apps.}
\label{tab:tp-per-app}
\begin{tabular}{lrr}
\toprule
\textbf{Undeclared third parties} & \textbf{N} & \textbf{\%} \\
\midrule
1                  & 61  & 37.0\% \\
2                  & 36  & 21.8\% \\
3                  & 26  & 15.7\% \\
4                  & 12  & 7.3\%  \\
5 or more          & 30  & 18.2\%  \\
\midrule
Total              & 165 & 100.0\% \\
\midrule
\multicolumn{3}{l}{\textit{Mean} 2.78 \quad \textit{Median} 2 \quad \textit{Range} 1--14} \\
\bottomrule
\end{tabular}
\end{table}

\subsection{RQ2: How many Mini Apps use Telegram’s Standard Bot Privacy Policy, and does exposing a custom one correspond to better compliance?}

\label{sec:rq2}
As shown in Table~\ref{tab:pp_type}, out of 278 working Telegram Mini Apps, 219 (78.8\%) rely on the Standard Bot Privacy Policy and do not expose a custom privacy policy. Only 59 Mini Apps (21.2\%) provided a custom one.
\begin{table}[ht]
\centering
\caption{Privacy policy type among working Mini Apps. Apps are split between Telegram's Standard Bot Privacy Policy, assigned by default when the developer does not provide one, and a Custom Privacy Policy explicitly exposed by the developer. Percentages are computed over the 278 working Mini Apps.}
\label{tab:pp_type}
\begin{tabular}{lcc}
\toprule
\textbf{Privacy Policy type} & \textbf{N} & \textbf{\%} \\
\midrule
Standard (Telegram default) & 219 & 78.8\% \\
Custom & 59 & 21.2\% \\
\bottomrule
\end{tabular}
\end{table}

These results indicate that the vast majority of Mini Apps rely on Telegram's default policy, while only a minority explicitly define and expose their own privacy policy.
With respect to the 219 Mini Apps relying on the Standard Bot Privacy Policy, Table~\ref{tab:pp_violations} shows that 132 of them (60.3\%) exhibit at least one privacy policy violation, i.e.\ they contact at runtime at least one third party that is not mentioned in the applicable policy. Among the 59 Mini Apps with a custom privacy policy, 33 (55.9\%) exhibit at least one violation. To establish whether this apparent gap reflects a genuine difference in disclosure accuracy rather than sampling variation, we tested the two groups for independence using Pearson's $\chi^2$~\cite{Pearson1992} without continuity correction, as the expected cell frequencies were sufficiently large (the smallest expected count is 23.98), and, as a confirmatory analysis, Fisher's exact test~\cite{fisher1922interpretation}. Neither test reveal a significant association between policy type and violation status ($\chi^2(1) = 0.36$, $p = 0.55$; Fisher's exact test $p = 0.55$; odds ratio $= 1.20$, Cram\'er's $V = 0.04$). 

Although Mini Apps with a custom privacy policy show a slightly lower violation rate, the difference is not statistically significant. The observed effect is small ($h \approx 0.088$), and the estimated power for detecting an effect of this magnitude is only about 0.09 at $\alpha = 0.05$. Thus, the current sample is unlikely to detect such a small effect reliably.

A sensitivity analysis indicates that an effect of approximately $h = 0.41$, corresponding to a difference of about 20 percentage points, would be needed to achieve 80\% power with the current sample sizes. Therefore, these results do not provide evidence that policy type affects the violation rate, but they cannot exclude the possibility of a small genuine effect.
%\mc{si riesce a calcolare la Power di questo test? questo serve per sapere se non c'e' differenza significativa, o se hai troppi pochi punti sperimentali per apprezzare una differenza significativa}\lf{fixed}\lv{rivista da controllare}

\begin{tcolorbox}[colback=gray!5,colframe=blue!40, boxrule=0.3mm, arc=0mm, left=1mm, right=1mm, top=1mm, bottom=1mm]
\textbf{RQ2: How many Mini Apps use Telegram's Standard Bot Privacy Policy, and does exposing a custom one correspond to better compliance?}
The large majority of working Mini Apps (78.8\%, 219/278) rely on Telegram's default Standard Bot Privacy Policy rather than exposing a custom one, suggesting that most Mini Apps rely on the platform-provided disclosure rather than an application-specific policy.
%developers do not engage directly with privacy disclosure and instead default to the platform's generic template. 
Exposing a tailored Privacy Policy, however, does not correspond to a statistically distinguishable improvement in disclosure accuracy. Violations affect 60.3\% (132/219) of apps under the standard policy and 55.9\%(33/59) of those with a custom one, a difference of 4.4 percentage points that is not significant ($p = 0.55$). 
%Although our limited sample size prevents us from ruling out a small genuine effect, 
Although our sample does not provide sufficient power to reliably detect an effect as small as the one observed, the practical takeaway remains unchanged: more than half of the custom policies still fail to mention at least one third party contacted during our observation.
\end{tcolorbox}

\begin{table}[h]
\centering
\caption{Privacy violations by privacy policy type. For each policy type, the table reports the number of Mini Apps adopting it and how many exhibit at least one violation, i.e.\ contact at runtime a third party not mentioned in the applicable policy. Percentages are computed row-wise, over the Mini Apps of that policy type; the last row gives the overall rate across the 278 working Mini Apps.}
\label{tab:pp_violations}

\begin{tabular}{lccc}
\toprule
\textbf{Privacy Policy type} & \textbf{Total} & \textbf{With violation} & \textbf{\%} \\
\midrule
Standard (Telegram default) & 219 & 132 & 60.3\% \\
Custom & 59 & 33 & 55.9\% \\
\midrule
\textbf{Total} & \textbf{278} & \textbf{165} & \textbf{59.4\%} \\
\bottomrule
\end{tabular}%

\end{table}
\subsection{RQ3: At which stages of the Mini App lifecycle do privacy violations occur?}
\label{sec:rq3}

Among the 165 working apps found to violate their stated privacy policy, we further disaggregated violations by the lifecycle stage at which they occur: during app opening, only during use, or at both stages. This distinguishes stage-specific violations from those observed across both monitored stages.

\begin{table}[h]
\centering
\caption{Breakdown of violations by session phase: whether undeclared
third-party contacts occur only at opening, only during use, or across both.
Percentages are over the 165 violating Mini~Apps and do not sum to 100 due to
rounding.}
\label{tab:violation_phase}
\begin{tabular}{p{4.5cm}cc}
\toprule
\textbf{Sub-category} & \textbf{N} & \textbf{\%} \\
\midrule
Violation only at opening  & 53 & 32.1\% \\
Violation only during use & 24 & 14.5\% \\
Violation at both opening and during use & 88 & 53.3\% \\
\midrule
\textbf{Total with at least one violation} & \textbf{165} & \textbf{100.0\%} \\
\bottomrule
\end{tabular}
\end{table}

As shown in Table~\ref{tab:violation_phase}, more than half of the violating apps (88 out of 165, 53.3\%) exhibit undeclared communication with third parties at \emph{both} stages, while violations confined to a single stage are more than twice as frequent at opening (53, 32.1\%) than during use (24, 14.5\%). This is a particularly relevant finding: privacy issues are strongly front-loaded and frequently observed across both stages, rather than confined to a single, isolated event. 
The large majority of violating apps (141 out of 165, 85.5\%) already exhibit undeclared third-party communication during Mini App opening, and in almost two-thirds of these cases (88 out of 141, 62.4\%), undeclared communication is also observed during subsequent user interaction. 
This pattern is consistent with the architecture of Telegram Mini Apps, which are web applications loaded in an embedded browser: initialization scripts, third-party libraries, and tracking code may be executed at page load, while additional behaviors can be triggered later by user interactions, navigation between views, and asynchronous callbacks. 
The prevalence of both-stage violations (31.7\% of all
working apps) suggests that undeclared data sharing is not limited to the bootstrap phase, but often spans multiple stages of Mini App execution. 
%but a structural characteristic of the app's data collection design. 
Conversely, violations that surface only after the user starts interacting with the app are the least common case (24 apps, 14.5\%), which indicates that observing the launch alone would already identify most Mini Apps in which we observe a violation, whereas the reverse does not hold.%which indicates that observing the launch alone would already expose most of the undeclared traffic, whereas the reverse does not hold.

Taken together, the two categories involving the opening stage account for 141 of the 165 violating apps (85.5\%), i.e.\ the large majority of violations already manifest before any meaningful in-app interaction.%any user action is possible. 
This timing is significant in light of a second, complementary observation: across all 278 working Mini Apps, manual inspection of the launch screen revealed no instance of a consent banner, dialog, or equivalent mechanism. In no case was the user given an in-app opportunity to accept, refuse, or configure data processing before it occurred. The transmissions observed at the opening stage therefore precede not merely the user's first in-app interaction, but any observable in-app mechanism for expressing a preference.

In the specific case of Telegram, this timing is particularly relevant to the platform's own acceptance model. Under the Standard Bot Privacy Policy, continued access to and use of a Mini App is construed as acceptance of the applicable policy. Our measurements nevertheless show that undeclared third-party communication may occur immediately at page load, before the user can perform any interaction within the Mini App or encounter an application-level consent mechanism. This creates a temporal tension between the immediate execution of data flows and a policy model in which acceptance is inferred from continued access and use. 

\begin{tcolorbox}[colback=gray!5,colframe=blue!40, boxrule=0.3mm, arc=0mm, left=1mm, right=1mm, top=1mm, bottom=1mm]
\textbf{RQ3: At which stages of the Mini App lifecycle do privacy violations occur?}

% Privacy violations tend to be persistent rather than confined to a single moment.
% More than half of the violating apps (53.3\%, 88/165) exhibit undeclared
% third-party communication both at opening and during use, while single-stage
% violations occur more than twice as often at opening (32.1\%) than during actual use
% (14.5\%). This recurring pattern suggests that undeclared data sharing is embedded in
% the app's initialization and overall data collection design, rather than being a
% one-off or transient technical artifact.
% Moreover, none of the 278 working apps presented any consent mechanism at launch, so
% the 85.5\% of violating apps, whose misbehavior already begins at the opening stage,
% leave the user no opportunity to mediate.
Privacy violations are frequently observed across multiple stages of Mini App execution rather than being confined to a single moment. More than half of the violating apps (53.3\%, 88/165) exhibit undeclared third-party communication both at opening and during use, while single-stage violations occur more than twice as often at opening (32.1\%) than during actual use (14.5\%). This recurring pattern suggests that undeclared data sharing is not limited to isolated or transient events, but often spans both initialization and subsequent user interaction. Moreover, none of the 278 working apps presented an observable consent mechanism at launch, while 85.5\% of violating apps already contacted undeclared third parties during the opening stage, before any meaningful in-app interaction or opportunity to express a preference.

\end{tcolorbox}

\subsection{RQ4: What types of user data are predominantly shared with undeclared third parties?}% \lv{versione più corretta di quello che misuriamo. se ti piace cambiala anche dalle altre parti}
\label{sec:rq4}
Beyond identifying \emph{whether} and \emph{when} a violation occurs, it is equally
relevant to identify
\emph{what} categories of user data are involved. During manual
inspection, each detected violation was annotated with the type(s) of data observed in the
corresponding outbound request. Every violating Mini App contained at least one identifiable
data category, so the annotation covers the full set of 165 violating apps.
Table~\ref{tab:data_types} reports the distribution of data categories across them.

\begin{table}[h]
\centering
\caption{Categories of user data transmitted to undisclosed third parties. Percentages are
over the 165 violating Mini Apps. Categories are not mutually exclusive.}
\label{tab:data_types}
\begin{tabular}{lcc}
\toprule
\textbf{Data type} & \textbf{N} & \textbf{\% of violating} \\
\midrule
Device info           & 157 & 95.2\% \\
User and Profile info &  63 & 38.2\% \\
\bottomrule
\end{tabular}
\end{table}

Device information is by far the most frequently transmitted category, appearing in all
but eight of the violating apps (157/165, 95.2\%). User and profile information, such as
Telegram specific identifiers and account metadata, is the second most common category
(63/165, 38.2\%). 
This is of particular concern, as such data can be directly associated with an identifiable Telegram account, rather than only with a device. %as such data is tied directly to a specific, identifiable individual rather than to a device. 
Since the two categories are not mutually exclusive and every violating app falls in at least one of them, the two
counts also determine their overlap: 55 apps (33.3\% of the violating ones, 19.8\% of all
working Mini Apps) transmit both categories to undisclosed recipients, 102 (61.8\%)
transmit device information only, and just 8 (4.8\%) transmit profile information without
any accompanying device attribute. 
In other words, one violating app in three transmits both device and account-related information to undeclared recipients.
%In other words, one violating app in three discloses a device fingerprint and an account identity to the same undeclared third parties.

Within the \emph{Device info} category, the transmitted attributes are not scattered
across apps: a set of five fields appears together in almost every request, namely, device
model, Android version, browser (WebView) version, device language and Telegram app version.
A second group of attributes, comprising device OS, screen resolution and OS version, appears
less frequently and predominantly after the user has started interacting with the app. Taken
in isolation, none of these fields is particularly sensitive, and each has a plausible
technical justification, such as layout adaptation or compatibility checks. 
%Their systematic co-occurrence, however, is what makes them problematic: already the five recurring fields,
% i.e.\ device model, Android version, WebView version, language and Telegram app version,
% jointly constitute a stable device fingerprint, sufficient to re-identify the same device
% across sessions and, more importantly, across different Mini Apps sharing the same
% third-party recipient~\cite{specter2025fingerprinting, bouhenniche2026exadprinter}.
% The attributes of the second group, when present, further narrow the resulting identifier.
% Since the five recurring fields are derived from the User-Agent string and from the
% initialization payload, they are transmitted at page load, before any user action, and therefore before any opportunity for consent.
Their systematic co-occurrence, however, may increase their fingerprinting potential: the five recurring fields, device model, Android version, WebView version, language, and Telegram app version, provide a combination of device characteristics that may contribute to distinguishing or correlating devices across sessions and applications sharing the same third-party recipient~\cite{specter2025fingerprinting,bouhenniche2026exadprinter}.
The attributes of the second group, when present, may further increase the distinctiveness of this combination. Since the five recurring fields are derived from the User-Agent string and the initialization payload, they are already observable in requests generated at page load, before any meaningful user action and therefore before any observable in-app opportunity to express a preference.

The \emph{User and Profile info} category exhibits a different pattern. 
Rather than a
heterogeneous set of attributes, what is forwarded is primarily information contained in the Telegram initialization
payload (\texttt{initData}), often relayed to third parties with little or no filtering. In
decreasing order of frequency, the observed fields are: Telegram user ID, first and last name,
\texttt{initData} signature, authentication timestamp, profile picture URL, chat instance and
chat type, \texttt{initData} hash, the entire \texttt{initData} blob, username, and language
code.

\begin{tcolorbox}[colback=gray!5,colframe=blue!40, boxrule=0.3mm, arc=0mm, left=1mm, right=1mm, top=1mm, bottom=1mm]
\textbf{RQ4: What types of user data are predominantly shared with undeclared third parties?}

Device information is involved in almost every violating app (95.2\%, 157/165), followed by
User and Profile data (38.2\%, 63/165); one violating app in three (33.3\%) transmits both categories to undeclared recipients.%discloses both.
% Undeclared sharing is therefore not limited to low-sensitivity technical parameters: the
% device attributes are transmitted as a stable co-occurring set that amounts to a device
% fingerprint, sent at page load and before any opportunity for consent, while the
% user-related fields consist largely of the Telegram \texttt{initData} payload relayed with
% little filtering.
Undeclared sharing is therefore not limited to technical parameters: device attributes are transmitted as a recurring co-occurring set with potential fingerprinting value, often at page load and before any observable in-app opportunity to express a preference, while user-related fields consist largely of information contained in the Telegram \texttt{initData} payload relayed with little filtering.
\end{tcolorbox}

\section{Discussion}
\label{sec:discussion}
%\lv{eviterei di dire che l’assenza di banner costituisce automaticamente una violazione dell’Art. 5(3) ePrivacy: il consenso è richiesto per storage/access non strettamente necessario, ma esistono eccezioni; inoltre il vostro traffico include anche comunicazioni che potrebbero non ricadere tutte nello stesso modo sotto l’Art. 5(3).eviterei “none of the analyzed Mini Apps is in a position to claim a valid legal basis”: il GDPR ammette più basi giuridiche, non solo il consenso.modificherei fortemente “the platform's own theory of consent is therefore violated on its own terms”: Telegram dice che il continued access/use costituisce acceptance della policy, ma questa acceptance contrattuale non equivale necessariamente al consenso privacy richiesto dalla normativa.toglierei completamente l’idea che tApps Center sia un ambiente curato da Telegram: nel resto del paper lo avete correttamente definito community-driven.starei attento a “Telegram is the only party that could verify Mini App behavior at scale”: meglio “is uniquely positioned” o “is in a particularly strong position”. LA SEZIONE SOTTO è rifatta con questa considerazione}
The results presented in Section~\ref{sec:exp} quantify the divergence between the data practices declared by Telegram Mini Apps and those observed at runtime. RQ1 shows that 59.4\% of the working Mini Apps in our dataset contact at least one third-party domain that is not disclosed in the privacy policy applicable to the app (Section~\ref{sec:rq1}); RQ2 shows that 78.8\% of the analyzed apps rely on Telegram's generic Standard Bot Privacy Policy instead of publishing an app-specific one, and that this policy does not enumerate specific third-party recipients (Section~\ref{sec:rq2}); RQ3 shows that more than half of the violating apps (53.3\%) exhibit undeclared third-party communication both at opening and during use (Section~\ref{sec:rq3}); and RQ4 shows that the data transmitted to undeclared third parties predominantly include device attributes (95.2\% of the violating apps) and Telegram profile information (38.2\%), with 33.3\% of the apps transmitting both categories (Section~\ref{sec:rq4}).

\textbf{The absence of observable consent mechanisms.}
A complementary finding concerns not \emph{what} is disclosed but \emph{whether users are given an opportunity to express a preference before data processing occurs}. Across the entire set of 278 working Mini Apps we analyzed, \textbf{not a single one} presented a cookie banner, consent dialog, or other interactive mechanism through which the user could accept, refuse, or granularly configure data processing at launch.

This observation is particularly relevant under Article~5(3) of the ePrivacy Directive~\cite{eprivacy}, which generally requires prior consent when information is stored on or accessed from a user's terminal equipment, except where such access is strictly necessary for transmitting a communication or providing a service explicitly requested by the user. Accordingly, where the third-party tracking technologies we observe involve non-essential access to or storage of information on the terminal equipment, the absence of a prior consent mechanism raises a potential compliance concern.

As reported in Section~\ref{sec:rq3}, 85.5\% of violating apps (141/165) already contact undeclared third parties during the opening stage, i.e., before any meaningful in-app interaction. This finding is especially relevant for analytics and advertising services, for which comparable Web deployments commonly rely on a consent-management mechanism when their operation requires consent under applicable law.

The absence of an application-level consent interface must also be considered in light of Telegram's own acceptance model. The Standard Bot Privacy Policy~\cite{MiniAppPP}, which governs 78.8\% of the apps in our sample, states that continued access to and use of a Third-Party Service constitutes acceptance of the policy, the Telegram Bot Terms, and the Telegram Mini App Terms. The Mini App Terms similarly construe continued access and use as acceptance of the applicable terms. This contractual acceptance mechanism, however, is distinct from any consent that may be required for a specific data-processing operation under applicable privacy law. Our measurements further show that undeclared third-party communications can occur immediately at page load, before users can interact with the Mini App or encounter any in-app mechanism through which processing choices could be expressed. This creates a temporal and transparency gap between the execution of data flows and the user's practical ability to understand or control them.

The consequence is therefore not that every observed Mini App necessarily lacks a valid legal basis for every processing operation, but that the ecosystem provides users with very limited observable control over third-party data flows. Merely making a privacy policy accessible from the bot profile does not itself constitute consent to processing that legally requires it, and the default Standard Bot Privacy Policy may apply without the Mini App presenting an application-specific privacy choice at launch.

\textbf{Implications for the platform.}
We argue that the gap documented by our measurements cannot be addressed solely at the level of individual developers, and that Telegram is particularly well positioned to introduce platform-level safeguards. Our sample was drawn from tApps Center, a widely used community-driven catalogue of Telegram Mini Apps, and 59.4\% of the analyzed applications contact at least one third party not disclosed by the applicable privacy policy. Regardless of how applications are discovered, Telegram is the common execution environment through which these Mini Apps are launched and is therefore in a privileged position to observe or constrain their runtime behavior at scale. We identify three concrete mechanisms that could strengthen this ecosystem.

\begin{itemize}

\item \emph{Runtime verification of declared data flows.} Telegram could complement existing platform controls with a dynamic check similar to the one described in Section~\ref{sec:methodology}: launching the Mini App, observing its outbound traffic, and comparing contacted domains with the recipients or categories disclosed in the applicable privacy policy. Our results suggest that runtime verification is practically useful because undeclared communications recur across multiple applications and frequently involve a limited set of analytics and advertising infrastructures.

\item \emph{Controls against hot updates.} Unlike packaged mini apps, Telegram Mini Apps are Web applications served from developer-controlled infrastructure (Section~\ref{sec:back}). Their content can therefore change after any initial review without requiring redistribution of a package through Telegram. A Mini App may consequently exhibit different runtime behavior after it has first been inspected. One-off vetting can therefore provide only a point-in-time view of the application. Mitigating this limitation would require periodic or event-triggered re-verification and could be complemented by platform-level restrictions on external origins, for example through an enforceable Content Security Policy or an equivalent allow-list mechanism associated with the Mini App.

\item \emph{Verification that apps remain functional.} Of the 442 entries drawn from tApps Center, 151 (34.2\%) corresponded to bots or frontends that were no longer functional: 124 had unreachable frontends and 27 bots no longer existed. Besides reducing catalogue quality, stale entries may introduce additional security risks. In particular, a domain that expires while remaining referenced by an existing bot or catalogue entry could potentially be re-registered and used to serve content different from that originally associated with the Mini App. Periodic liveness checks and removal or flagging of unreachable applications would reduce this exposure.

\end{itemize}

Taken together, these measures highlight an asymmetry between developer responsibility and platform-level observability. Telegram's Terms place responsibility for the operation of Mini Apps and the handling of user data on the corresponding third-party Service Providers~\cite{TermsBot,TermsMiniApp}. The Mini App Terms further state that, after data have been transmitted to a Service Provider, Telegram does not have access to or control over the subsequent exchange between the user and that provider. At the same time, Telegram explicitly recognizes Mini Apps as independently operated third-party services and reserves the ability to restrict access to individual Service Providers or modify how users access the Mini App platform.

Our findings show that relying exclusively on developer-side disclosure leaves an important observability gap: users cannot readily determine which external recipients receive their data, while external authorities cannot continuously inspect each application individually. Telegram, as the common execution platform, is therefore in a particularly strong position to introduce scalable runtime transparency and verification mechanisms, even where ultimate responsibility for the Mini App's processing practices remains with the third-party developer.

\section{Threats to Validity and Limitations}
\label{sec:Limitation}
%\lv{modifiche fatte per evitare claim troppo forti sulla rappresentatività statistica, rendere più prudente il passaggio sul consenso/GDPR, e distinguere meglio tra threats to validity e tool limitations.}
%We discuss here the factors that may affect the validity of our findings, following the standard classification into construct, internal, and external validity, and we conclude with the practical limitations of our tool.

\subsection{Construct Validity}

% \textbf{Unvalidated data extractors.} The detection of sensitive information in outbound traffic relies on a set of regular-expression extractors applied to normalized request contents (Section~\ref{sec:implementation}). We did not establish a manually labelled ground truth against which to measure the precision and recall of these extractors. Consequently, both false positives (e.g., a benign string matching an identifier pattern) and false negatives (e.g., values encoded in formats not covered by our normalization pipeline) may propagate into the figures reported in Section~\ref{sec:exp}. We mitigated this risk by recursively decoding nested payloads, by requiring a minimum value length to filter trivial matches, and by manually reviewing the extracted values for every app included in the manual annotation stage; nonetheless, the accuracy of the extraction stage remains unquantified, and the reported violation counts should be read as approximations rather than exact measurements.

\textbf{Single Telegram account.} Our methodology requires an active, logged-in Telegram account on the analysis device, since the Start Bot Module (Section~\ref{sec:methodology}) operates within the Telegram client itself. All measurements were therefore collected under a fixed account configuration, including a specific language tag, premium subscription status, theme parameters, and profile metadata. Since these attributes are part of the initialization payload transmitted to Mini Apps, the specific data categories observed in RQ4 may not generalize to accounts configured differently: an app might transmit additional or different attributes depending on the profile it receives. The presence of undeclared third-party communication, however, does not depend on account configuration, so this threat concerns the characterization of \emph{what} is shared rather than \emph{whether} sharing occurs.

\textbf{Scope of the consent-mechanism inspection.} The absence of any consent artifact reported in Section~\ref{sec:rq3} was established through manual inspection of the launch screen of each of the 278 working Mini Apps, and through the sessions observed during our analysis. It therefore attests that no consent banner, dialog, or equivalent control was presented at launch in the executions we observed, but it does not exclude the possibility that such a mechanism exists along application paths not reached by our exploration, for instance behind authentication, in a settings menu, or in views triggered by interaction sequences outside the explored budget. 

We note, however, that this limitation does not affect the timing observation developed in Section~\ref{sec:discussion}: 85.5\% of the violating apps contact undeclared third parties at page load, before any meaningful in-app interaction and before any observable in-app opportunity to express a preference. Where a processing operation relies on consent as its legal basis, GDPR consent must satisfy the conditions set out in Articles~6 and~7~\cite{gdpr6,gdpr7}; our analysis does not determine the legal basis applicable to each individual transmission.

\subsection{Internal Validity}

\textbf{Manual annotation of custom policies.} For the 59 Mini Apps exposing a custom Privacy Policy, the classification of a third-party contact as \emph{declared} or \emph{undeclared} rests on a manual reading of the policy text and its comparison against the observed traffic (Section~\ref{sec:methodology}). This judgment is inherently interpretive: Privacy Policies frequently refer to recipients in generic terms, such as ``analytics providers'' or ``our partners'', without naming the specific domains contacted at runtime, and deciding whether such a formulation covers a given endpoint involves a degree of discretion. We adopted a conservative criterion, counting a recipient as declared whenever the policy could plausibly be read as encompassing it, which biases our results toward under-reporting violations. 

\textbf{Exclusion of platform hosts.} Requests to hosts in our \texttt{EXCLUDED\_HOSTS} set are filtered out before third-party inspection (Section~\ref{sec:implementation}). We built this set by manually inspecting 100 randomly sampled Mini Apps and keeping only those hosts that recurred across unrelated apps, under a trust model for Google platform services. If a Mini App were to transmit privacy-relevant data through one of these excluded hosts, our pipeline would not detect it, which again biases our results toward under-reporting.%If a Mini App exfiltrates data through one of these channels, our pipeline would not detect it, which again biases our results toward under-reporting.

\textbf{Residual infrastructure traffic.}
Our pipeline separates first-party from Third Party communication by inferring
the Mini App domain as described in the Traffic Analysis Module (Section~\ref{sec:implementation}) and
by filtering the hosts listed in

\texttt{EXCLUDED\_HOSTS}. Neither mechanism can
capture the full set of endpoints that legitimately belong to a Mini App's own
infrastructure: developers routinely serve assets and backend APIs through cloud and CDN providers (e.g., Cloudflare, Microsoft Azure) under domains that cannot be listed beforehand and have no syntactic connection to the first-party origin. Requests of this kind are consequently retained in the report even when they carry first-party traffic routed through a delegated provider, generating noise in the Traffic Analysis Module results. Our subsequent policy-level review mitigates this issue when the role of the contacted service can be established, but some residual misclassification may remain.

\subsection{External Validity}
\label{sec:extValidity}

\textbf{Sample representativeness and temporal snapshot.} Our study relies on a random sample of 278 working Mini Apps drawn from a population of 991 entries in \textit{DtApps}. Following standard practices in empirical software engineering~\cite{ferrari2026evaluatingllmsobfuscationdetection,alecci2025damflow,alecci2026taskflow}, this sample size was computed to ensure a 95\% confidence level with a 5\% margin of error. These parameters provide statistical coverage with respect to the tApps Center sampling frame, but do not guarantee representativeness of the entire Telegram Mini App ecosystem, since applications not listed in tApps Center fall outside our population. Furthermore, because Mini Apps are web applications dynamically served from developer-controlled infrastructure, our measurements inherently represent a temporal snapshot. Developers can alter their applications at any time without platform intervention. Consequently, an app found compliant today might introduce undeclared tracking tomorrow, meaning that longitudinal replication would be required to establish whether the documented patterns remain stable over time

% \mc{qui devi menzionare il fatto che non hai studiato tutte le mini app, ma una rappresentanza, dicendo come ti sei sincerato di prendere una "buona" rappresentanza}\lf{fixed}

\subsection{Limitations}

\textbf{Unreachable Mini Apps due to missing UI entry points.} The Start Bot
Module (Section~\ref{sec:methodology}) opens a Mini App by navigating to the
associated bot profile and selecting the ``Open App'' shortcut. However, not every bot exposes this shortcut directly from its profile page: in some cases, the Mini App can only be launched through an inline button embedded in a specific chat message, a deep link shared outside Telegram, or a menu configured differently by the developer. When the ``Open App'' entry point is absent from the profile, our automated pipeline is unable to reach the Mini App, and the corresponding bot is excluded from the analysis, as was the case for 13 of the 442 entries drawn during sampling (Section~\ref{sec:exp}). These apps may still be accessible through alternative launch mechanisms, so their exclusion is a coverage limitation of our tooling rather than a property of the ecosystem. It introduces a potential sampling bias toward Mini Apps that follow the most common, profile-exposed launch pattern, while apps relying on alternative or non-standard entry points remain outside the scope of our current tool.

\textbf{Exploration coverage.} The Mini App Interaction Module (Section~\ref{sec:methodology}) explores each app by clicking randomly, within a fixed budget of 240 seconds per run and $k = 5$ runs per app. Neither the budget nor the randomized strategy guarantees that all execution paths are reached: functionality behind authentication, multi-step flows requiring a specific click sequence, and views reachable only from deeper navigation states may remain unvisited, and any undeclared third-party contact triggered exclusively along those paths goes unobserved. 

This limitation primarily affects the interaction stage. The initialization stage requires no exploratory interaction; the Mini App is simply launched and observed during page load. Consequently, the finding that 85.5\% of violating apps already exhibit undeclared third-party communication at opening (Section~\ref{sec:rq3}) does not depend on interaction-path coverage, although it may still be affected by other sources of measurement error discussed above.

\section{Conclusion and Future works}
\label{sec:conclusion}

This paper presented the first empirical study of Privacy Policy compliance in Telegram Mini Apps. Applying our dynamic-analysis framework TeleGapper to 278 working apps drawn from tApps Center, we found that 59.4\% contact at least one third party that the applicable privacy policy does not disclose (RQ1); that the large majority of Mini Apps rely on Telegram's generic Standard Bot Privacy Policy (RQ2, 78.8\%), while apps that do provide a custom policy exhibit a statistically indistinguishable violation rate; that undeclared third-party communication frequently spans multiple stages of execution (RQ3, 53.3\% at both monitored stages); and that the data involved extend beyond technical device attributes to identifiable user and profile information in 38.2\% of the violating apps, largely in the form of information contained in the Telegram \texttt{initData} payload relayed to third parties with little or no filtering (RQ4). Across all 278 apps, not one presented an observable consent mechanism at launch, while 85.5\% of violating apps contact undisclosed third parties before any meaningful in-app interaction is possible.

Taken together, these results reveal a substantial gap between the privacy disclosures presented to users and the runtime behavior of Telegram Mini Apps. In particular, reliance on the default privacy framework provides limited transparency into the third-party communications performed by individual Mini Apps, while undeclared data sharing frequently begins during application initialization. Since our measurements cover only Mini Apps sampled from tApps Center and rely on deliberately conservative classification criteria, the figures we report should be interpreted as estimates for our sampling frame and may understate the prevalence of undeclared data flows.

Several directions remain open. On the tooling side, we plan to extend the framework to iOS, which requires a different automation and interception stack, and to investigate root-free TLS interception in order to lower the barrier to independent replication. On the measurement side, a longitudinal campaign would establish whether the violations we document persist across app updates, a question made particularly relevant by the hot-update capability discussed in Section~\ref{sec:discussion}: apps served from developer-controlled infrastructure can change behavior without redistribution through the platform. Finally, we intend to extend the analysis beyond privacy compliance toward the detection of security vulnerabilities and malicious behavior in Mini Apps, complementing the methodology presented here with a security-oriented analysis.

\section*{Funding}
This work was partially supported by project SERICS (PE00000014) under the NRRP MUR program funded by the EU - NGEU.

\section*{Data and code availability}
Released artefacts include (i) the \emph{Mini App Scraper} described in Section~\ref{sec:implementation}, the Selenium-based crawler for the tApps Center, together with $D_{\mathrm{tApps}}$, the resulting dataset of 991 catalogue entries collected in the first half of June 2026 (bot name, app title, category); (ii) \textsc{TeleGapper} itself, comprising the Appium-based \emph{Start Bot} and \emph{Dynamic Analysis} Modules, the proxy configuration
used for traffic interception, and the \emph{Traffic Analyzer Module}, including the full set of regular-expression extractors and the \texttt{EXCLUDED\_HOSTS} list applied before third-party inspection.

We do not release the network traces, the per-app reports (\textit{e1}, \textit{e2}), the interaction logs, or the manual annotations produced during the campaign. Two distinct concerns motivate this decision. First, the traces record the initialization payload of the Telegram account used to drive the analysis, including its user identifier, \texttt{initData} signature and
authentication timestamps; these are personal data of the authors, and they cannot be redacted without removing the very fields on which the extraction pipeline operates. Second, the per-app artifacts attribute specific undeclared data flows to named Mini Apps and to the companies operating them.

Code and derived data are publicly available at \url{https://github.com/Mobile-IoT-Security-Lab/TeleGapper}. Raw third-party data remain subject to the original providers’ terms.

\section*{Responsible Disclosure}
In alignment with ethical research practices in cybersecurity and privacy, we are currently undertaking a responsible disclosure process. We are collecting the available contact information (e.g., support emails, official Telegram channels) for the developers of the Mini Apps found in violation of their privacy policies to notify them of our findings. Furthermore, because we identify these privacy gaps as a platform-level issue, we are simultaneously sharing our results and proposed countermeasures with Telegram’s security and privacy team.

\section*{CRediT authorship contribution statement}
\textbf{Luca Ferrari}: Writing – original draft, Software, Methodology, Data curation, Conceptualization, Validation, Investigation, Formal Analysis, Project administration, Writing – review \& editing. \textbf{Luca Verderame}: Methodology, Writing – review \& editing, Supervision.\textbf{ Mariano Ceccato}: Methodology, Writing – review \& editing, Supervision.

\section*{Declaration of competing interest}
The authors declare that they have no known competing financial interests or personal relationships that could have appeared to influence the work reported in this paper.

\section*{Declaration of generative AI use}
During the preparation of this work, the authors used ChatGPT and Grammarly for grammar checking, spelling correction, paraphrasing, and language revision. The authors subsequently reviewed and edited all generated suggestions and take full responsibility for the content of the paper.

%-------------------------------------------------------------------------------

\bibliographystyle{plain}
\bibliography{bib}

\end{document}